\documentclass[11pt]{article}
\usepackage[a4paper,margin=0.88in]{geometry}
\usepackage{amsmath,amssymb,bm}
\usepackage{graphicx}
\usepackage{booktabs}
\usepackage{array}
\usepackage{siunitx}
\DeclareSIUnit\mmHg{mmHg}
\usepackage{microtype}
\usepackage{hyperref}
\usepackage{natbib}
\usepackage{authblk}
\usepackage{xcolor}
\usepackage{enumitem}
\usepackage{float}
\title{\textbf{Inverse-Designed Anisotropic Intrastromal Corneal Exoskeletons for Keratoconus: A Three-Dimensional Computational Feasibility Study}}

\author[1]{Juan Sumaya-Mart\'inez}
\author[2]{Alan Altamirano-Torres}
\affil[1]{Faculty of Sciences, Universidad Aut\'onoma del Estado de M\'exico, Toluca 50000, M\'exico}
\affil[2]{Centro de Microcirug\'ia Ocular, Atlacomulco, Estado de M\'exico, M\'exico}
\date{\today}

\begin{document}
\sloppy
\maketitle

\begin{abstract}
Keratoconus combines localized stromal thinning and biomechanical weakening with strongly asymmetric optical aberrations, whereas current intrastromal implants remain predominantly arc- or ring-based. We propose G-EXO, an inverse-designed anisotropic intrastromal corneal exoskeleton: a thin, sectorial scaffold whose position, angular span, stromal depth, and effective stiffness are optimized against a coupled biomechanical--optical objective. The central hypothesis is that spatially targeted intrastromal reinforcement can redistribute the pressure-driven deformation field so as to reduce an asymmetric optical observable, such as coma, without requiring a globally symmetric flattening response.

The study was developed in stages. Reduced nonlinear screening was first used to explore the design space and revealed that reinforcement need not coincide with the ectatic apex. We then constructed an explicit ribbon-lattice representation and a reproducible three-dimensional tetrahedral finite-element (FEM) screening model with spatially weakened ectatic tissue, physiological intraocular pressure (IOP), and a bonded intrastromal reinforcement domain. In the current linearized FEM, a candidate sector centered near $-155^\circ$, spanning $60^\circ$, occupying approximately $0.55$--$0.85$ of local stromal depth, and having an effective tangent-stiffness ratio $E_s/E_c\approx3$ reduced pressure-induced anterior-surface coma over the tested IOP range. The optical effect was mesh-sensitive: at 2188 tetrahedra, interpolation of the deformed anterior surface onto a dense pupil grid before Zernike fitting yielded a representative coma reduction of approximately $4.8\%$, whereas displacement and stress metrics were substantially more stable.

These results do not establish clinical efficacy. Rather, they demonstrate computational feasibility, expose the numerical requirements of optical post-processing on deformable corneal meshes, and identify a testable design principle: non-ring, spatially targeted intrastromal reinforcement can be optimized directly against asymmetric corneal deformation. Patient-specific tomography, nonlinear anisotropic hyperelasticity, calibrated scaffold mechanics, mesh-converged optical analysis, and ex-vivo validation are required before translational interpretation.
\end{abstract}

\section{Introduction}

Keratoconus (KC) is an ectatic corneal disorder characterized by localized thinning, biomechanical weakening, progressive or stable anterior protrusion, irregular astigmatism, and higher-order optical aberrations. The biomechanical and optical manifestations are intrinsically coupled: a spatially heterogeneous reduction in stromal stiffness changes the response of the cornea to intraocular pressure (IOP), which in turn modifies anterior and posterior surface curvature and the corresponding wavefront error \citep{andreassen1980,gefene2009,han2013,falgayrettes2023}. The resulting optical degradation is typically asymmetric, with coma being particularly important in decentered cones.

Conventional management ranges from optical rehabilitation with rigid or scleral contact lenses to corneal cross-linking (CXL), intracorneal ring segments (ICRS), corneal allogenic intrastromal ring segments (CAIRS), and keratoplasty in advanced cases. ICRS and CAIRS are especially relevant to the present work because they demonstrate that intrastromal implantation can mechanically redistribute corneal shape. Finite-element studies have shown that the geometry, depth, material properties, and placement of ring segments affect both local stress and postoperative curvature \citep{kling2013,lago2015,bagheri2021,garcia2023}. Recent clinical CAIRS studies further show that biologic intrastromal material can produce substantial tomographic and visual changes and can reduce coma in selected cases \citep{yesilirmak2026,gunn2026,barbara2026}.

Despite these advances, the design space remains strongly constrained by ring or arc geometry. This is a natural consequence of implant history and surgical practicality, but it is not a requirement imposed by biomechanics. A decentered ectatic cornea is neither axisymmetric nor mechanically homogeneous. Therefore, a mechanically optimal reinforcement field need not be circumferentially symmetric and need not be centered on the ectatic apex.

This motivates the present question:

\begin{quote}
Can a thin, spatially selective, anisotropic intrastromal scaffold be inverse-designed so that the pressure-driven deformation field of an ectatic cornea is modified specifically to reduce asymmetric optical aberrations?
\end{quote}

We call the proposed device class \emph{G-EXO}: an inverse-designed intrastromal corneal exoskeleton. The term ``exoskeleton'' is used here in a mechanical sense: the scaffold is conceived as a localized load-redistributing architecture embedded within the stromal volume. It is not intended to imply rigid fixation or a clinically validated implant.

The present paper has four goals. First, we formulate the optical--biomechanical inverse-design problem mathematically. Second, we define a reproducible 3-D FEM screening model that can evaluate sectorial intrastromal reinforcement. Third, we analyze the numerical sensitivity of a wavefront-relevant quantity, namely Zernike coma, to structural mesh resolution and surface interpolation. Fourth, we establish a roadmap from synthetic proof-of-concept modeling toward patient-specific nonlinear FEM and ex-vivo validation.

\section{Relation to prior intrastromal and biomechanical models}

\subsection{Finite-element models of keratoconus}

Corneal FEM has evolved from isotropic and orthotropic approximations toward nonlinear anisotropic descriptions that explicitly account for collagen fiber organization and regional heterogeneity. Early models demonstrated that localized thinning and local reductions in elastic or shear modulus can generate keratoconus-like deformation under IOP \citep{gefene2009}. Three-dimensional anisotropic hyperelastic models later showed that both local stiffness distribution and stromal layer mechanics affect displacement and refractive power \citep{han2013}. Patient-specific modeling has also been used to predict geometric changes after ICRS implantation \citep{lago2015}.

More sophisticated constitutive frameworks incorporate collagen architecture through Holzapfel--Gasser--Ogden (HGO) or related strain-energy formulations. Such models can represent fiber recruitment, anisotropy, and regional differences in stromal architecture \citep{whitford2018,falgayrettes2023}. These developments are essential for quantitative prediction, but they also increase parameter-identification requirements. In the present work, we intentionally separate \emph{screening} from \emph{validation}: the executed 3-D FEM uses a spatially varying linearized tangent modulus, while HGO/GOH is specified as the required next-stage constitutive model.

\subsection{Intrastromal ring segments and patient-specific design}

ICRS alter corneal shape through a combination of geometric displacement, arc-shortening, and stress redistribution. FEM studies have shown that segment geometry and implantation strategy change the predicted postoperative shape \citep{kling2013,bagheri2021,garcia2023}. Lago et al.\ demonstrated patient-specific FEM of ICRS implantation using measured geometry and a hyperelastic corneal model \citep{lago2015}. These works establish a crucial precedent: intrastromal implants can and should be treated as patient-specific biomechanical design problems.

CAIRS extend this concept using allogenic stromal tissue rather than a synthetic ring material. Recent studies report improvements in keratometry, visual acuity, and corneal regularity, including coma reduction in some advanced KC cohorts \citep{yesilirmak2026,barbara2026}. However, published CAIRS approaches still use ring-like segments and clinical nomograms rather than free-form inverse optimization.

\subsection{Design gap addressed by G-EXO}

The distinction between G-EXO and established intrastromal-segment approaches is therefore not primarily material choice but \emph{design freedom}. In G-EXO, the reinforcement is treated as a spatial field or scaffold whose position, angular extent, radial extent, depth, and directional stiffness are variables of an optimization problem.

\begin{table}[H]
\centering
\caption{Conceptual comparison of established intrastromal-segment approaches and the proposed G-EXO design space.}
\begin{tabular}{p{0.19\linewidth}p{0.24\linewidth}p{0.24\linewidth}p{0.25\linewidth}}
\toprule
 & ICRS & CAIRS & G-EXO (this work) \\
\midrule
Typical geometry & Synthetic arc/ring segment & Allogenic stromal arc segment & Free-form anisotropic sector or lattice \\
Design variables & Arc, thickness, optical zone, depth & Arc, width, thickness, depth, axis & Position, span, radii, depth, stiffness tensor, microstructure \\
Primary target & Keratometry / refractive regularization & Clinical regularization / nomogram & Explicit optical--stress--volume objective \\
Material & Synthetic polymer & Allogenic corneal tissue & To be selected/calibrated \\
Current status & Clinical & Clinical / emerging & Computational feasibility \\
\bottomrule
\end{tabular}
\label{tab:comparison}
\end{table}

\section{G-EXO hypothesis and inverse-design formulation}

\subsection{Mechanical hypothesis}

Let the undeformed cornea occupy a domain $\Omega_c$ and the scaffold a subdomain $\Omega_s(\mathbf p)\subset\Omega_c$. The key hypothesis is that a local increase in directional tangent stiffness changes the displacement field
\begin{equation}
\mathbf u(\mathbf x;\mathbf p)
\end{equation}
under IOP in a way that can reduce a chosen optical observable. Because the ectatic geometry is asymmetric, the mechanically useful reinforcement need not lie directly over the weakest point. Reinforcement on a contralateral or oblique sector can alter the global compatibility conditions of the shell and redistribute meridional/circumferential strain.

This is conceptually different from seeking a uniform flattening effect. G-EXO instead seeks a displacement field whose low-order and high-order optical consequences are favorable.

\begin{figure}[H]
\centering
\includegraphics[width=0.64\linewidth]{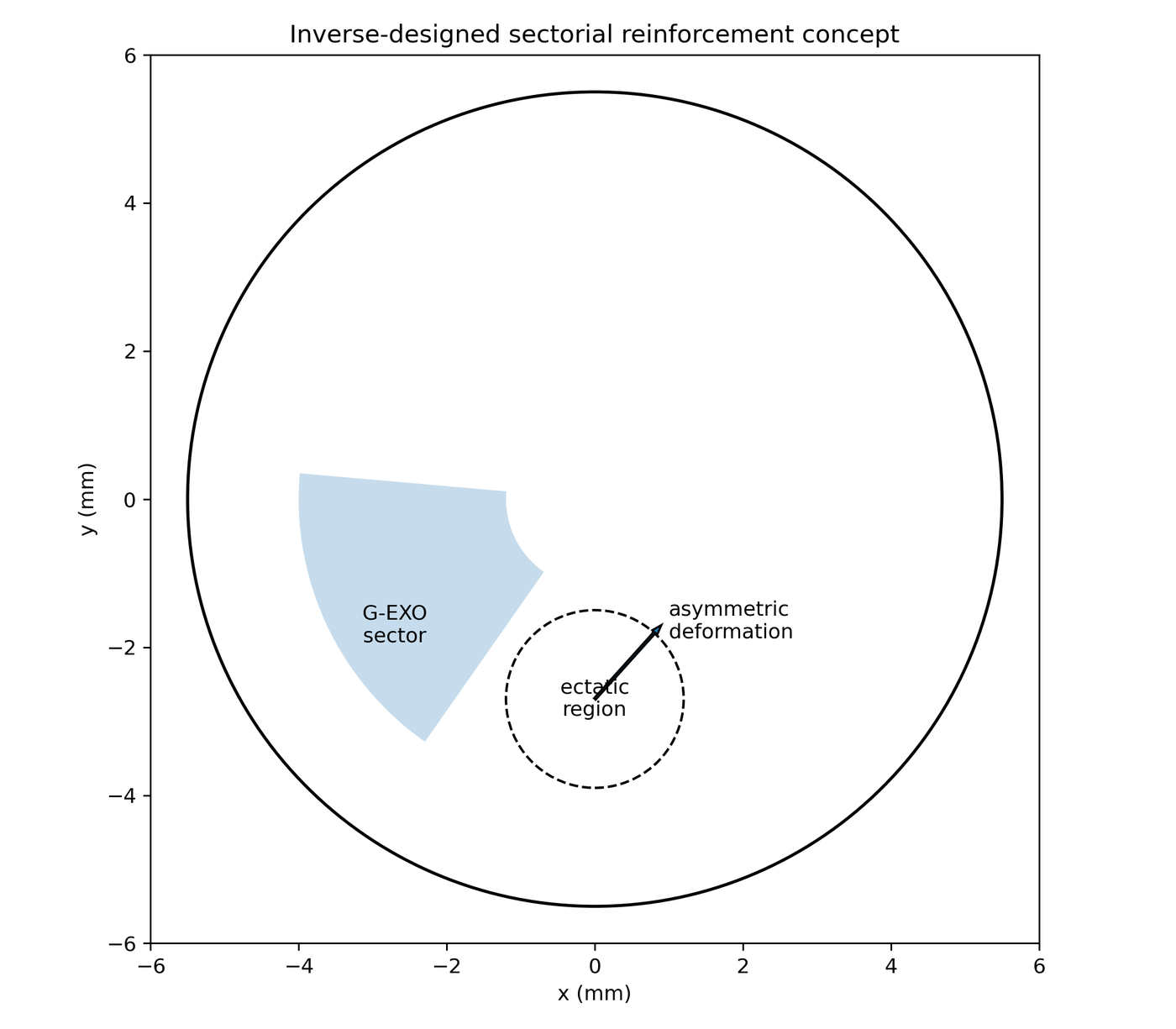}
\caption{Schematic of the G-EXO concept. A sectorial intrastromal reinforcement redistributes the deformation field generated by IOP. The illustration is conceptual and not a surgical map.}
\label{fig:concept}
\end{figure}

\subsection{Design vector}

A minimal macroscopic design vector is
\begin{equation}
\mathbf p=
\left(
\theta_c,\Delta\theta,r_i,r_o,z_{\min},z_{\max},
\gamma_s,\mathbf a
\right),
\end{equation}
where
\begin{itemize}[nosep]
\item $\theta_c$ is the angular center of the scaffold,
\item $\Delta\theta$ is its angular span,
\item $r_i$ and $r_o$ are inner and outer radii,
\item $z_{\min}$ and $z_{\max}$ are normalized stromal-depth limits,
\item $\gamma_s=E_s/E_c$ is an effective stiffness ratio, and
\item $\mathbf a$ represents microstructural anisotropy parameters.
\end{itemize}

A resolved lattice would expand $\mathbf a$ to include ribbon width, thickness, node reinforcement, cell angle, and orientation. In the present 3-D FEM, the scaffold is treated as an effective bonded domain.

\subsection{Optical--biomechanical objective}

The general objective is
\begin{equation}
J(\mathbf p)=
w_c\,C_{\mathrm{coma}}^2
+w_h\,\mathrm{HOA}_{\mathrm{RMS}}^2
+w_\sigma
\left(
\frac{\sigma_{95}}{\sigma_{\mathrm{ref}}}
\right)^2
+w_v
\left(
\frac{V_s}{V_{\mathrm{ref}}}
\right)
+w_r\,\mathcal R(\mathbf p),
\label{eq:objective}
\end{equation}
where
\begin{equation}
C_{\mathrm{coma}}
=
\sqrt{C_{3,-1}^2+C_{3,1}^2},
\end{equation}
$\sigma_{95}$ is a robust high-stress metric, $V_s$ is scaffold volume, and $\mathcal R$ is a regularization term penalizing impractical geometry, abrupt stiffness gradients, or minimum-feature violations.

The use of a stress percentile rather than only the maximum stress is intentional: pointwise maxima can be mesh-sensitive near geometric discontinuities, whereas a percentile is more robust during early design screening.

\begin{figure}[H]
\centering
\includegraphics[width=0.96\linewidth]{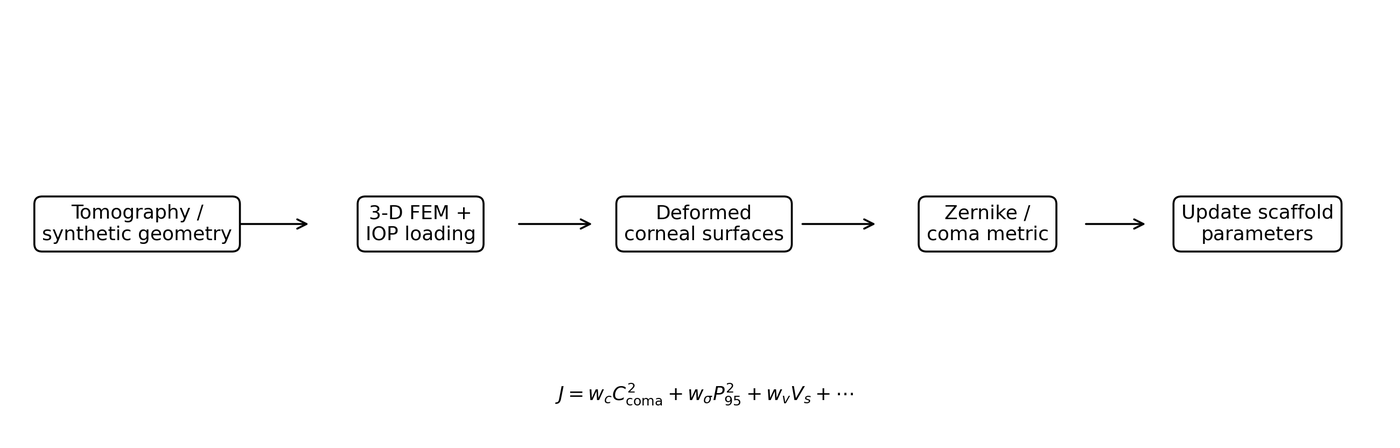}
\caption{Proposed inverse-design loop. The present work executes the geometry $\rightarrow$ FEM $\rightarrow$ surface displacement $\rightarrow$ Zernike portions of the workflow and performs a limited design search.}
\label{fig:workflow}
\end{figure}

\begin{figure}[H]
\centering
\includegraphics[width=0.72\linewidth]{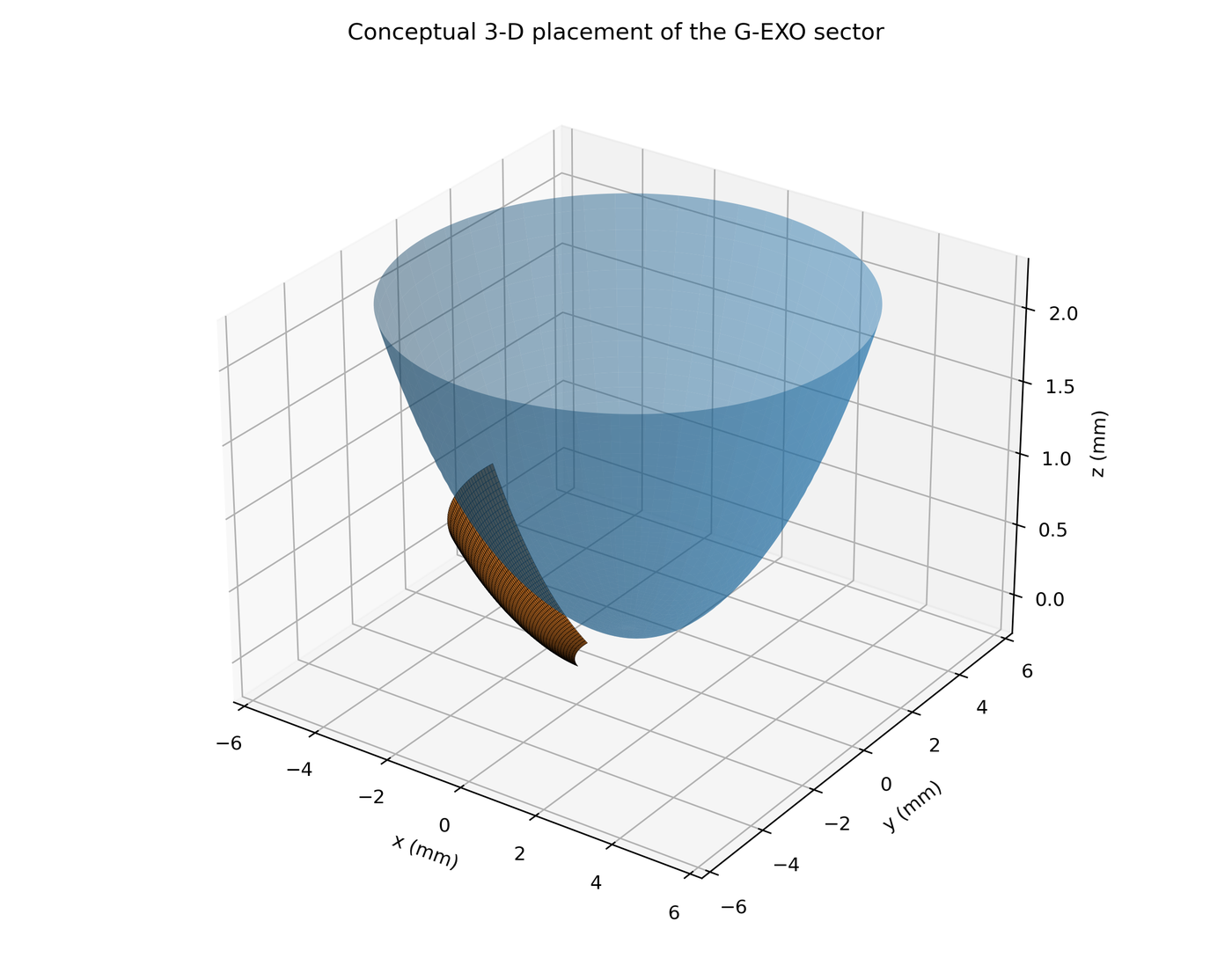}
\caption{Conceptual three-dimensional placement of the sectorial G-EXO within the corneal volume. The scaffold is shown below the anterior surface for visualization. The geometry is illustrative and not a patient-specific surgical plan.}
\label{fig:3dconcept}
\end{figure}

\section{Synthetic ectatic geometry}

\subsection{Reference corneal shape}

The screening cornea is represented by an \SI{11}{\milli\metre} diameter domain with a nominal anterior radius
\begin{equation}
R_c=\SI{7.8}{\milli\metre}.
\end{equation}
The baseline anterior sag is approximated by
\begin{equation}
z_{\mathrm{ant},0}(r)=
R_c-\sqrt{R_c^2-r^2}.
\end{equation}

To introduce a decentered ectatic perturbation, a Gaussian field
\begin{equation}
g(x,y)=
\exp\left[
-\frac{(x-x_c)^2}{2s_x^2}
-\frac{(y-y_c)^2}{2s_y^2}
\right]
\end{equation}
is used, with center
\begin{equation}
(x_c,y_c)\approx(0.41,-2.95)\ \mathrm{mm}.
\end{equation}
The anterior surface becomes
\begin{equation}
z_{\mathrm{ant}}(x,y)
=
z_{\mathrm{ant},0}(r)
+A_z\,g(x,y),
\end{equation}
where $A_z$ is a synthetic ectatic elevation amplitude.

The local stromal thickness is
\begin{equation}
t(x,y)=t_0-\Delta t\,g(x,y),
\end{equation}
with
\begin{equation}
t_0=\SI{0.545}{\milli\metre},
\qquad
\Delta t=\SI{0.080}{\milli\metre}.
\end{equation}
The posterior surface is then defined approximately by
\begin{equation}
z_{\mathrm{post}}(x,y)=z_{\mathrm{ant}}(x,y)-t(x,y).
\end{equation}

These values generate an asymmetric test geometry. They are not claimed to be a patient-specific reconstruction.

\subsection{Why a synthetic model is useful}

A synthetic geometry permits controlled hypothesis testing before patient-specific inverse design. It also makes the numerical workflow reproducible. However, the quantitative output of such a model cannot be mapped directly onto a clinical case. The role of the synthetic model is therefore to answer a narrower question: \emph{does a localized reinforcement domain have the capacity to change a coma-related deformation mode in the expected direction?}

\section{Three-dimensional FEM formulation}

\subsection{Executed screening constitutive model}

The executed 3-D model uses small-strain, linearized isotropic elasticity as a tangent approximation:
\begin{equation}
\bm{\sigma}
=
\lambda\,\mathrm{tr}(\bm{\varepsilon})\mathbf I
+2\mu\bm{\varepsilon},
\end{equation}
with
\begin{equation}
\bm{\varepsilon}
=
\frac{1}{2}
\left(
\nabla\mathbf u+\nabla\mathbf u^T
\right),
\end{equation}
and Lam\'e constants
\begin{equation}
\lambda
=
\frac{E\nu}{(1+\nu)(1-2\nu)},
\qquad
\mu
=
\frac{E}{2(1+\nu)}.
\end{equation}

A nearly incompressible value $\nu=0.45$ is used for screening. The baseline tangent modulus is
\begin{equation}
E_0=\SI{0.30}{\mega\pascal}.
\end{equation}
Local ectatic weakening is represented by
\begin{equation}
E_c(x,y)
=
E_0
\left[
1-\eta\,g(x,y)
\right],
\qquad
\eta=0.45.
\label{eq:weakening}
\end{equation}

Equation~\eqref{eq:weakening} is an effective representation of tissue weakening; it does not distinguish matrix and collagen-fiber contributions.

\subsection{Scaffold domain}

Within the scaffold region $\Omega_s(\mathbf p)$, the effective tangent modulus is increased:
\begin{equation}
E_s(\mathbf x)
=
\max\left[
\gamma_s E_c(\mathbf x),E_{\min,s}
\right].
\end{equation}
For the candidate design reported below,
\begin{equation}
\gamma_s\approx3.
\end{equation}
Perfect bonding between scaffold and stroma is assumed. This is a deliberately strong interface assumption and should later be replaced by a cohesive or compliant-interface sensitivity study.

\subsection{Weak form}

The screening FEM solves the static equilibrium problem
\begin{equation}
\nabla\cdot\bm{\sigma}+\mathbf b=0
\quad
\mathrm{in}\ \Omega_c,
\end{equation}
subject to posterior pressure traction and peripheral constraints. The weak form is
\begin{equation}
\int_{\Omega_c}
\delta\bm{\varepsilon}:\bm{\sigma}\,dV
=
\int_{\Gamma_p}
\delta\mathbf u\cdot\mathbf t_p\,dA
+
\int_{\Omega_c}
\delta\mathbf u\cdot\mathbf b\,dV.
\end{equation}

Linear tetrahedral elements are used in the Python implementation. The element stiffness matrix is
\begin{equation}
\mathbf K_e
=
\int_{\Omega_e}
\mathbf B^T\mathbf D\mathbf B\,dV
\approx
\mathbf B^T\mathbf D\mathbf B\,V_e,
\end{equation}
because $\mathbf B$ is constant within a linear tetrahedron.

\subsection{Loading and boundary conditions}

IOP is applied to the posterior surface at
\begin{equation}
P\in\{10,15,20\}\ \mathrm{mmHg}.
\end{equation}
The initial screening model fixes a peripheral annulus approximately \SI{0.75}{\milli\metre} wide. This boundary condition is numerically convenient but mechanically restrictive. A publication-grade model should include a scleral extension or an elastic limbal foundation.

\section{Optical post-processing}

\subsection{Incremental surface wavefront surrogate}

The FEM provides an anterior-surface displacement field $u_z(x,y)$. In the present study, the optical metric is an \emph{incremental deformation metric}: Zernike coefficients are fitted to the pressure-induced anterior displacement over a 6-mm pupil. This is not equivalent to a full ocular wavefront or even a full corneal wavefront because it does not yet include posterior-surface refraction and ray tracing.

Let
\begin{equation}
u_z(\rho,\phi)
\approx
\sum_j a_j Z_j(\rho,\phi),
\end{equation}
where $Z_j$ are Zernike basis functions over the unit pupil. The coma magnitude is computed as
\begin{equation}
C_{\mathrm{coma}}
=
\sqrt{
a_{3,-1}^2+a_{3,1}^2
}.
\end{equation}

\subsection{Why surface interpolation is required}

A structural mesh is not automatically an optical sampling grid. If the pupil contains only a modest number of irregularly spaced surface nodes, direct least-squares Zernike fitting can be dominated by mesh geometry rather than physical deformation. Therefore, the refined workflow first interpolates the FEM displacement onto a dense, regular pupil grid and only then performs the Zernike decomposition.

This distinction proved essential in the present numerical audit.

\section{Design-search strategy}

\subsection{Reduced-order exploration}

Before running the tetrahedral FEM, reduced nonlinear shell models were used to explore candidate sectors, depths, and anisotropy. These models were computationally inexpensive and useful for identifying qualitative trends. They suggested that:

\begin{enumerate}[nosep]
\item reinforcement directly over the cone is not necessarily optimal,
\item contralateral or oblique reinforcement can reduce an asymmetric deformation mode,
\item stromal depth changes the mechanical leverage of the scaffold, and
\item optimizing optics and mechanics separately can lead to incompatible designs.
\end{enumerate}

The reduced models substantially overpredicted optical benefit relative to the later 3-D FEM. Their role in this paper is therefore hypothesis generation rather than quantitative prediction.

\subsection{Explicit lattice screening}

An explicit ribbon network was also explored to assess microstructural feasibility. The initial network contained approximately 60 nodes and 149 connecting segments. A corrected ribbon interpretation used nominal arm width near \SI{60}{\micro\metre}, local node widening toward \SI{100}{\micro\metre}, and a thin out-of-plane dimension in the computational geometry. Structural screening showed that local node reinforcement reduced stress concentration.

The lattice did not exhibit a robust negative effective Poisson ratio once coupled to an elastic stromal foundation. Consequently, the final concept is described as \emph{anisotropic}, not necessarily auxetic.

\subsection{Transition to 3-D FEM}

The local 3-D FEM search around the reduced-model candidate produced a best screening design with
\begin{align}
\theta_c &\approx -155^\circ,\\
\Delta\theta &\approx 60^\circ,\\
r_i &= \SI{1.2}{\milli\metre},\\
r_o &= \SI{4.0}{\milli\metre},\\
z/t &\in [0.55,0.85],\\
E_s/E_c &\approx 3.
\end{align}
The shift in optimum angle relative to earlier reduced-order screening emphasizes why the high-fidelity model must ultimately drive design.

\section{Results}

\subsection{Baseline and G-EXO response to IOP}

The screening FEM was evaluated at 10, 15, and \SI{20}{\mmHg}. In the initial mesh, the pressure-induced coma increased approximately linearly with IOP in both the untreated and reinforced models, while the candidate G-EXO sector consistently reduced the incremental coma relative to the same baseline model.

\begin{figure}[H]
\centering
\includegraphics[width=0.70\linewidth]{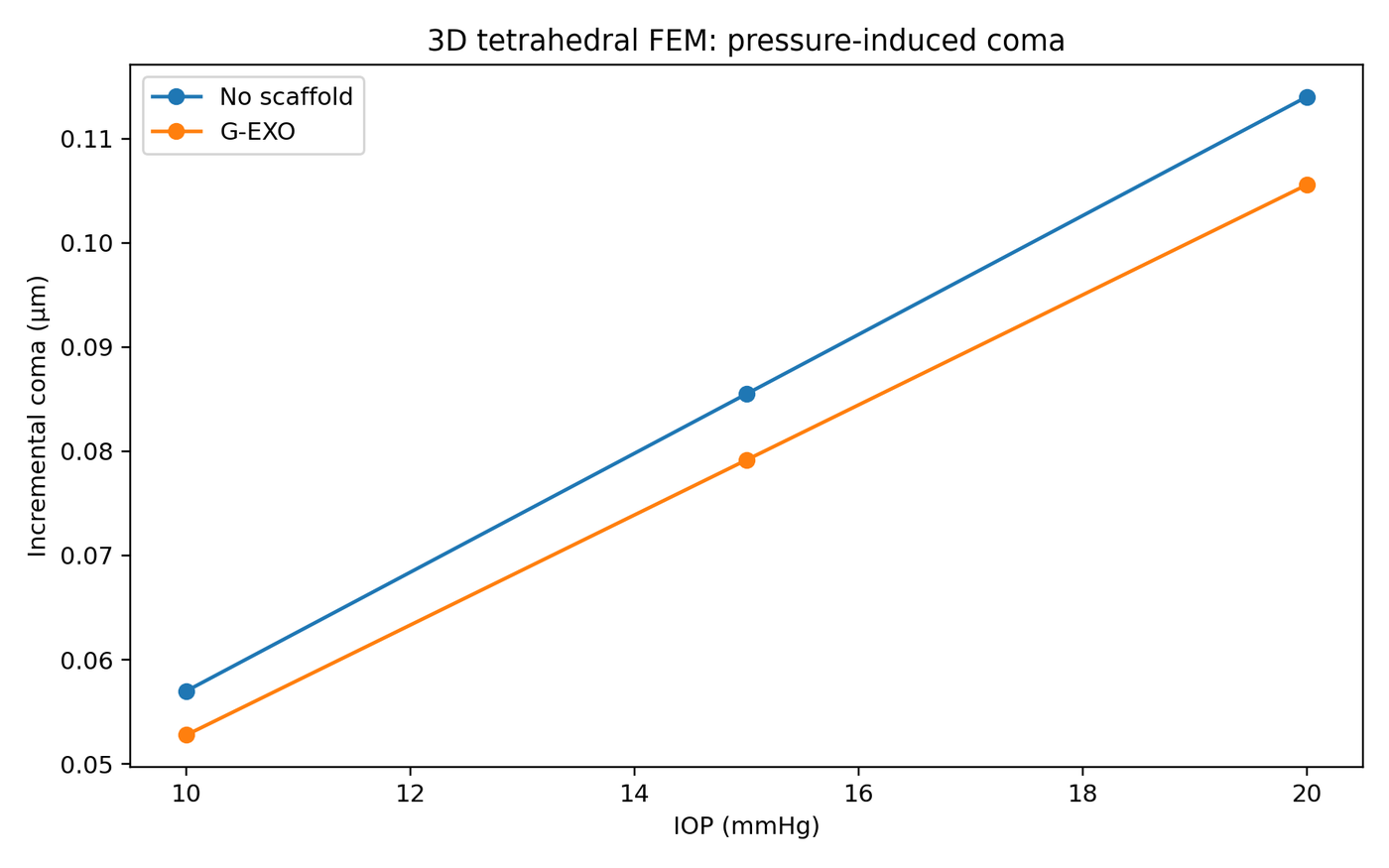}
\caption{Incremental anterior-surface coma versus IOP in the 3-D FEM screening model. This is a deformation-derived model metric rather than a clinical wavefront measurement.}
\label{fig:iop}
\end{figure}

The near-linear IOP dependence in the executed model is expected because the constitutive law is linearized. A nonlinear hyperelastic model should generally exhibit pressure-dependent tangent stiffness and therefore need not preserve the same proportionality.

\subsection{Mechanical displacement field}

At \SI{15}{\mmHg}, the maximum anterior displacement in the refined model remained close to \SI{5.3}{\micro\metre}. The scaffold changed the \emph{distribution} of deformation more than its global amplitude.

\begin{figure}[H]
\centering
\includegraphics[width=0.70\linewidth]{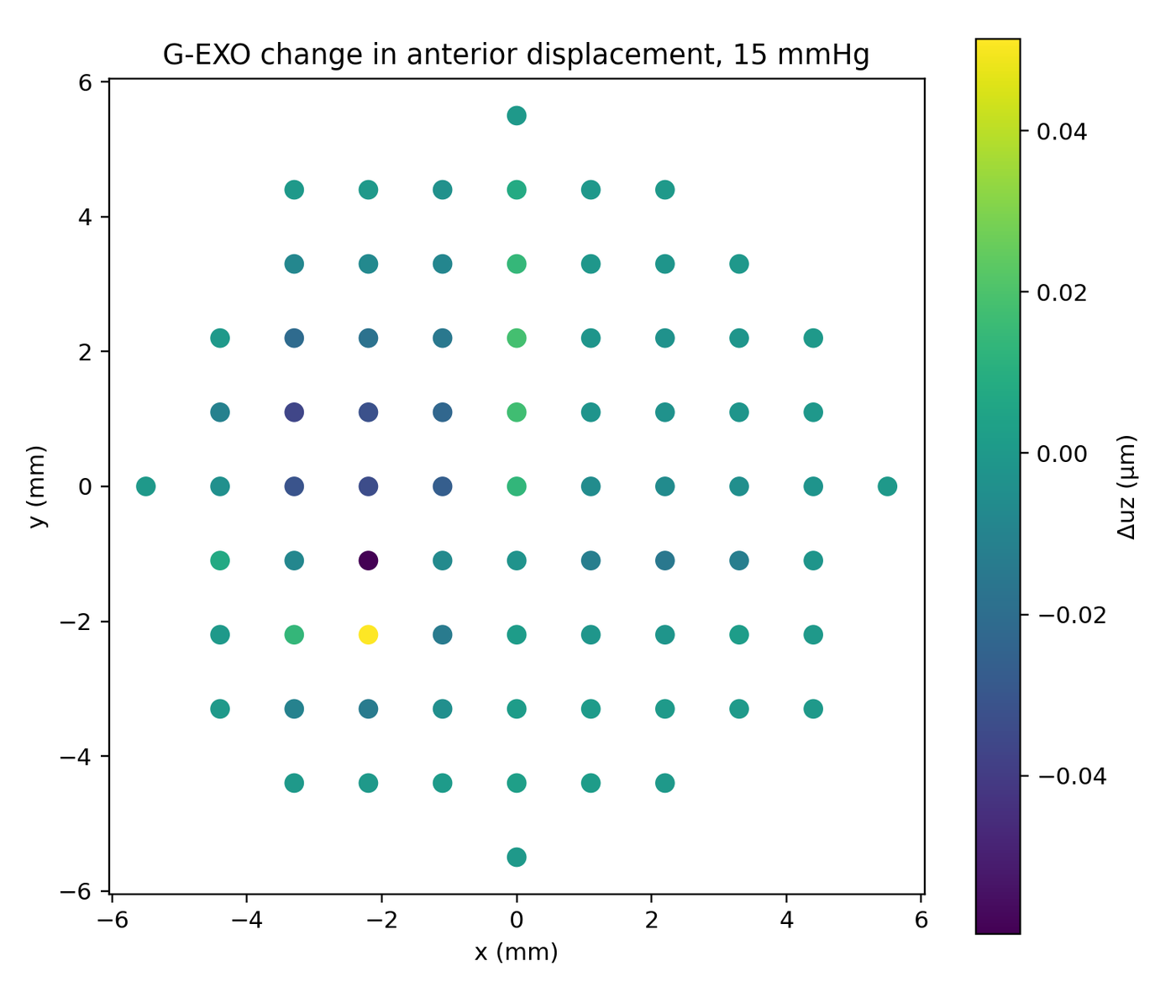}
\caption{Change in anterior displacement induced by the candidate G-EXO relative to the no-scaffold configuration at \SI{15}{\mmHg}.}
\label{fig:disp}
\end{figure}

This behavior is mechanically relevant. A successful G-EXO need not strongly reduce total displacement; it may instead redistribute deformation among azimuthal modes so that the coma-like component decreases.

\subsection{Stress field}

For the refined screening model, the 95th-percentile von Mises stress was approximately
\begin{equation}
\sigma_{95}\approx\SI{1.85}{\kilo\pascal}.
\end{equation}
The exact magnitude should not be interpreted physiologically because the constitutive model is simplified. More important is the relative numerical stability of the stress metric under mesh refinement.

\begin{figure}[H]
\centering
\includegraphics[width=0.70\linewidth]{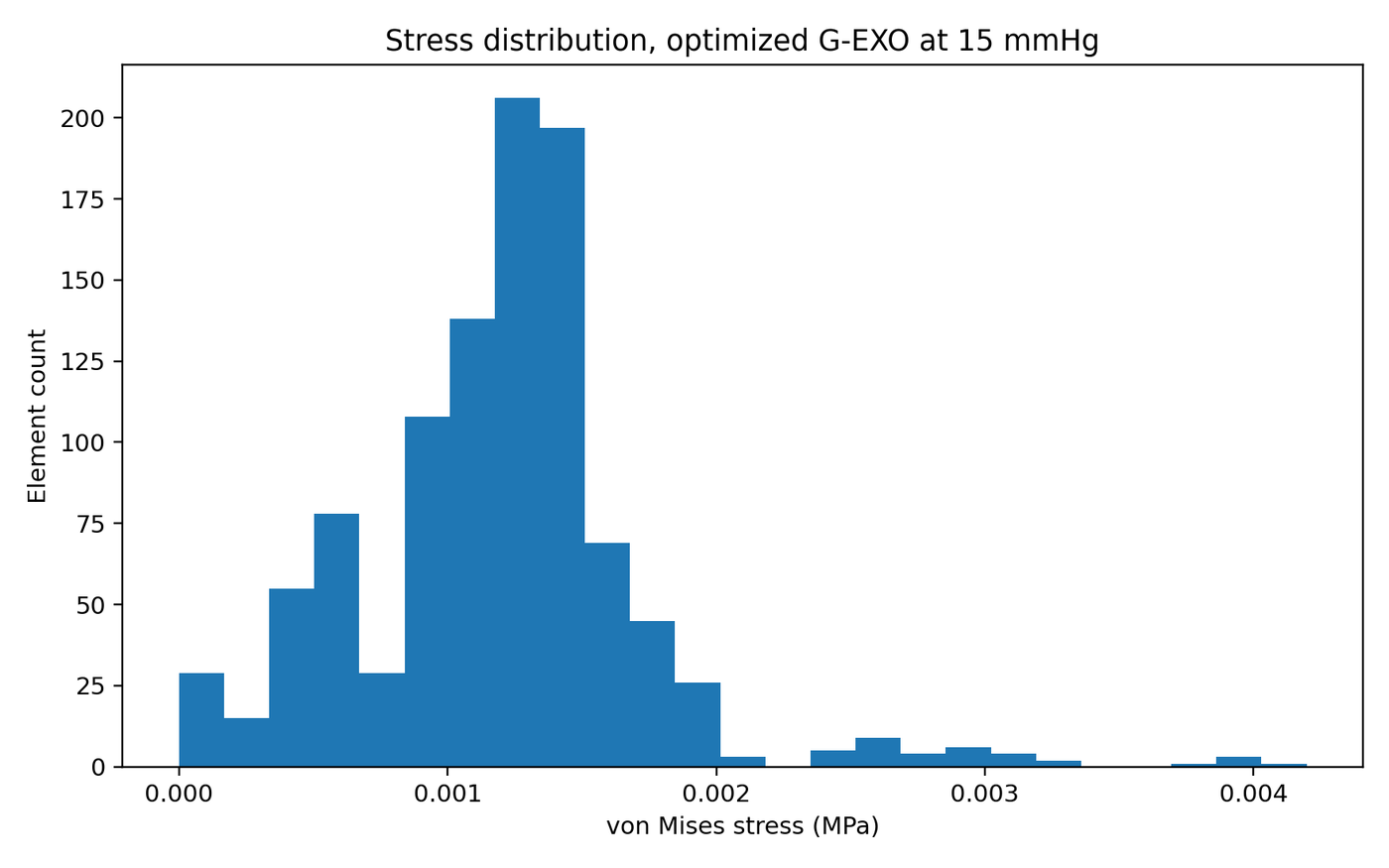}
\caption{Element-wise von Mises stress distribution for the optimized screening configuration at \SI{15}{\mmHg}.}
\label{fig:stress}
\end{figure}

\subsection{Optical convergence audit}

The most consequential numerical result was the difference between mechanical and optical convergence. A two-level comparison showed relatively small changes in displacement and stress, but substantial changes in directly fitted coma. Refinement to 2188 tetrahedra and interpolation of the deformed anterior surface before Zernike fitting reduced the sensitivity.

\begin{table}[H]
\centering
\caption{Representative optical convergence results at \SI{15}{\mmHg}.}
\begin{tabular}{lccc}
\toprule
Model & $C_{\rm base}$ ($\mu$m) & $C_{\rm GEXO}$ ($\mu$m) & Reduction \\
\midrule
1033 tetrahedra, nodal fit & 0.0751 & 0.0702 & 6.63\% \\
2188 tetrahedra, nodal fit & 0.1066 & 0.1023 & 4.04\% \\
2188 tetrahedra, interpolated fit & 0.0734 & 0.0699 & 4.83\% \\
\bottomrule
\end{tabular}
\label{tab:conv}
\end{table}

\begin{figure}[H]
\centering
\includegraphics[width=0.72\linewidth]{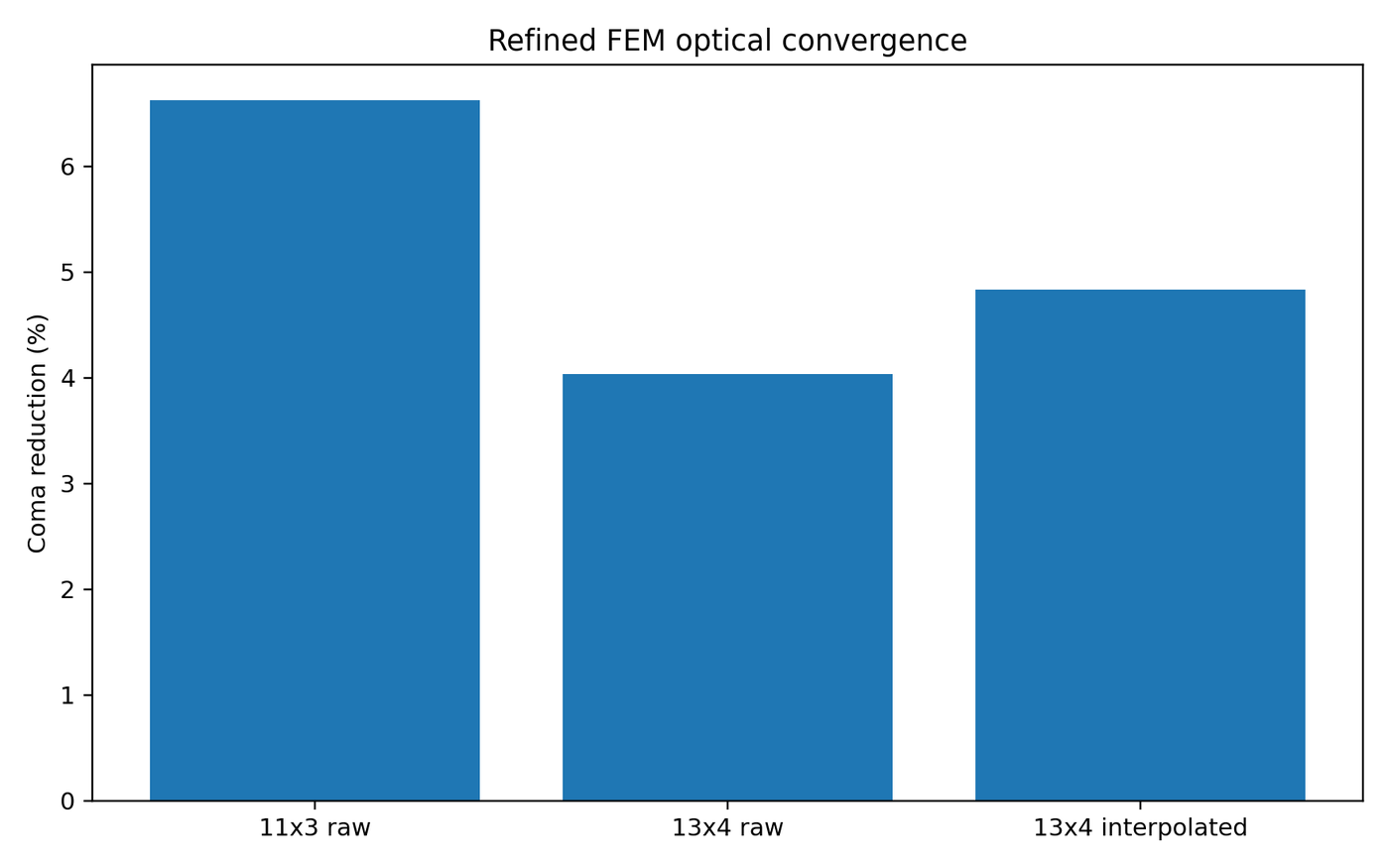}
\caption{Mesh and post-processing sensitivity of the coma reduction estimate. The interpolated 2188-tetrahedron result is used as the representative current estimate.}
\label{fig:conv}
\end{figure}

We therefore report the optical result as a \emph{trend near 5\%}, not as a converged therapeutic-effect size.

\subsection{Numerical interpretation}

The results support three distinct statements of different confidence:

\begin{enumerate}
\item \textbf{High confidence within the current model:} localized intrastromal stiffening changes the asymmetric displacement field.
\item \textbf{Moderate confidence within the current model:} the selected sector reduces the coma-like displacement component relative to the same baseline model.
\item \textbf{Low confidence for absolute prediction:} the exact percentage reduction cannot yet be interpreted clinically and is not fully mesh-converged optically.
\end{enumerate}

This hierarchy is important for avoiding overinterpretation.

\section{Why asymmetric reinforcement can reduce coma}

Coma is an odd-symmetry aberration. A spatially symmetric change in stiffness predominantly modifies symmetric deformation modes, while an off-axis or sectorial stiffness perturbation can couple more strongly to odd azimuthal modes.

Let the baseline tangent stiffness operator be $\mathcal K_0$ and the scaffold perturbation be $\delta\mathcal K(\mathbf p)$. The displacement under a load $\mathbf f$ satisfies
\begin{equation}
\left[
\mathcal K_0+\delta\mathcal K(\mathbf p)
\right]\mathbf u
=
\mathbf f.
\end{equation}
For a sufficiently small perturbation,
\begin{equation}
\delta\mathbf u
\approx
-\mathcal K_0^{-1}
\delta\mathcal K
\mathbf u_0.
\label{eq:perturb}
\end{equation}
Equation~\eqref{eq:perturb} makes the design principle explicit: a local stiffness perturbation produces a global displacement correction filtered by the compliance operator $\mathcal K_0^{-1}$. The useful scaffold position is therefore not necessarily where displacement is largest. It is where a stiffness perturbation most effectively projects onto the undesired optical mode.

If $\mathbf q_c$ is a discrete vector representing the coma-sensitive surface mode, then the first-order coma sensitivity is approximately
\begin{equation}
\delta C
\propto
-\mathbf q_c^T
\mathcal K_0^{-1}
\delta\mathcal K
\mathbf u_0.
\end{equation}
This expression provides a theoretical rationale for contralateral reinforcement and suggests a future adjoint-based optimizer.

\section{Proposed publication-grade hyperelastic model}

The current linearized FEM is appropriate for screening but not for quantitative corneal prediction. The next-stage COMSOL or open-source nonlinear model should include geometric nonlinearity, near incompressibility, and collagen-fiber anisotropy.

\subsection{Matrix and fiber energy}

A generic HGO/GOH-style strain energy can be written as
\begin{equation}
W
=
W_{\mathrm{iso}}(\bar I_1)
+
W_{\mathrm{vol}}(J)
+
\sum_{i=1}^{2}
W_{\mathrm{fib},i},
\end{equation}
with
\begin{equation}
W_{\mathrm{iso}}
=
C_{10}(\bar I_1-3),
\end{equation}
\begin{equation}
W_{\mathrm{vol}}
=
\frac{K}{2}(J-1)^2,
\end{equation}
and, for each preferred fiber family,
\begin{equation}
W_{\mathrm{fib},i}
=
\frac{k_1}{2k_2}
\left[
\exp\left(k_2\langle E_i\rangle_+^2\right)-1
\right].
\end{equation}

The parameters $C_{10}$, $k_1$, $k_2$, fiber dispersion, and local fiber orientation should be calibrated to experimental inflation/shear data or estimated through inverse FEM. Anisotropic constitutive models have previously demonstrated that local fiber architecture and stromal-layer mechanics materially affect keratoconic displacement and refractive behavior \citep{han2013,whitford2018,falgayrettes2023}.

\subsection{Ectatic weakening in a hyperelastic model}

Rather than reducing a single Young's modulus, the ectatic zone could be represented by spatial fields
\begin{equation}
C_{10}(\mathbf x),
\qquad
k_1(\mathbf x),
\qquad
\kappa(\mathbf x),
\end{equation}
allowing separate weakening of matrix, collagen recruitment, and fiber dispersion. Such a model can test whether G-EXO remains effective across plausible etiological mechanisms of KC.

\subsection{Scaffold--stroma interface}

Three interface models should be compared:
\begin{enumerate}[nosep]
\item perfect bonding,
\item thin compliant interphase, and
\item cohesive-zone or frictional contact if appropriate for the chosen material.
\end{enumerate}
The optical benefit should be considered robust only if it survives reasonable interface uncertainty.

\section{Patient-specific extension}

A true inverse-designed implant requires patient-specific anterior and posterior surfaces and pachymetry. The source package includes a pipeline that accepts a CSV containing
\begin{equation}
(x,y,z_{\mathrm{ant}},z_{\mathrm{post}})
\end{equation}
or anterior elevation plus local thickness.

The patient-specific workflow is:

\begin{enumerate}
\item import tomography;
\item remove invalid peripheral samples;
\item interpolate anterior and posterior surfaces onto a common grid;
\item generate through-thickness layers;
\item tetrahedralize the volume;
\item assign local stromal depth and fiber orientation;
\item solve the physiological preload;
\item introduce candidate G-EXO domains;
\item compute deformed surfaces and stress fields;
\item ray-trace or fit Zernike coefficients;
\item update the scaffold design.
\end{enumerate}

A major advantage of this formulation is that the objective can be changed without redefining the device concept. For one patient the target may be vertical coma; for another it may be a weighted combination of coma, trefoil, and irregular astigmatism.

\section{Potential scaffold materials and fabrication constraints}

This paper deliberately does not select a final scaffold material. The effective stiffness ratio in the current model is a design variable, not a material claim. A viable material must satisfy several constraints simultaneously:

\begin{enumerate}
\item optical transparency or minimal scattering,
\item biocompatibility,
\item mechanical stability at intrastromal hydration,
\item manufacturability at tens-of-micrometers feature scale,
\item compatibility with femtosecond-laser pocket insertion or another controlled placement method,
\item absence of cytotoxic degradation products, and
\item sufficient interface stability to transfer load without damaging adjacent lamellae.
\end{enumerate}

Bioengineered collagen-based or photocrosslinked hydrogel architectures are conceptually attractive because their mechanical properties can be tuned and because stromal analogues already exist in corneal tissue engineering. However, material choice requires independent experimental characterization and is beyond the evidence generated here.

\section{Experimental validation roadmap}

A translational validation sequence should proceed from mechanics to optics and only then to biological response.

\subsection{Bench mechanical characterization}

Candidate scaffold materials should undergo hydrated tensile/compression testing, cyclic loading, and stress-relaxation measurements. The quantities needed for FEM calibration include tangent modulus, nonlinear stiffening, hysteresis, swelling, and fatigue.

\subsection{Ex-vivo corneal inflation}

A practical next experiment is porcine or donor corneal inflation under controlled pressure. The protocol should compare:
\begin{enumerate}[nosep]
\item untreated cornea,
\item sham stromal pocket,
\item non-optimized symmetric scaffold,
\item inverse-designed sectorial scaffold.
\end{enumerate}
Digital image correlation, OCT, Scheimpflug imaging, or structured-light profilometry can provide surface displacement fields for direct comparison with FEM.

\subsection{Optical validation}

Topography/tomography before and after implantation should be converted to:
\begin{equation}
K_1,\ K_2,\ K_{\max},\ \mathrm{HOA}_{\mathrm{RMS}},
\ C_{3,-1},\ C_{3,1}.
\end{equation}
A strong validation would show that the FEM predicts not only flattening but the \emph{sign and magnitude of the targeted coma change}.

\subsection{Safety-focused outputs}

In addition to optical correction, experiments should monitor:
\begin{itemize}[nosep]
\item local stromal compression,
\item interface delamination,
\item keratocyte viability,
\item inflammatory response,
\item haze/scattering,
\item endothelial safety,
\item migration or rotation of the scaffold.
\end{itemize}

\section{Dimensionless interpretation of the G-EXO design space}

A useful way to generalize the present results beyond one synthetic geometry is to introduce dimensionless groups. Let $R_c$ be the characteristic corneal radius, $t_c$ the local corneal thickness, $E_c$ the local corneal tangent modulus, $E_s$ the effective scaffold modulus, and $P$ the IOP. We define
\begin{equation}
\Pi_P=\frac{PR_c}{E_ct_c},
\end{equation}
which compares pressure loading to local membrane stiffness, and
\begin{equation}
\Pi_E=\frac{E_s}{E_c},
\end{equation}
which measures scaffold stiffness contrast. Geometric scaffold parameters can be normalized as
\begin{equation}
\hat r_i=\frac{r_i}{R_c},
\qquad
\hat r_o=\frac{r_o}{R_c},
\qquad
\hat h_s=\frac{h_s}{t_c}.
\end{equation}

These groups clarify several design regimes. If $\Pi_E\rightarrow1$, the scaffold becomes mechanically indistinguishable from surrounding tissue and its effect vanishes. If $\Pi_E\gg1$, the reinforcement approaches a rigid inclusion and may introduce large stress gradients. Similarly, if $\hat h_s\ll1$, the scaffold behaves approximately as an embedded membrane or lattice; if $\hat h_s$ becomes comparable to stromal thickness, the problem becomes closer to bulk stromal replacement.

A future robust optimizer should therefore not optimize dimensional parameters independently. Instead, it should seek a region in the reduced space
\begin{equation}
\left(
\Pi_P,\Pi_E,\hat r_i,\hat r_o,z/t,\Delta\theta
\right)
\end{equation}
for which the optical benefit remains stable under uncertainty.

\section{Falsifiable predictions of the G-EXO hypothesis}

A useful computational concept should generate predictions that can be disproved. The G-EXO hypothesis makes several such predictions.

\subsection{Prediction 1: the optimum reinforcement need not overlap the cone apex}

If the mechanism is global load redistribution rather than local patching, the optimal scaffold sector should not necessarily coincide with the point of minimum thickness or maximum elevation. In a patient-specific inverse design, the optimum should instead track the spatial sensitivity of the coma mode to stiffness perturbation.

A direct experimental test is therefore to compare:
\begin{enumerate}[nosep]
\item a scaffold centered on the cone,
\item a scaffold placed at the inverse-designed location, and
\item a symmetric control scaffold of equal volume.
\end{enumerate}
If the cone-centered scaffold consistently performs as well as or better than the inverse-designed scaffold, the proposed contralateral-load-redistribution mechanism would be weakened.

\subsection{Prediction 2: equal-volume scaffolds can produce different optical outcomes}

The model predicts that geometry and orientation matter independently of implanted volume. Two scaffolds with the same volume but different angular placement should generate different coma changes. This distinguishes G-EXO from a simple ``more material gives more flattening'' mechanism.

\subsection{Prediction 3: coma can change without a large change in maximum displacement}

The current FEM already suggests that the scaffold can redistribute deformation while leaving the overall displacement scale nearly unchanged. Experimentally, this predicts that a measurable reduction in coma may occur even if apex displacement or Kmax changes only modestly.

\subsection{Prediction 4: optical benefit should depend on IOP}

Because the scaffold modifies a pressure-loaded structure, its effect should vary with IOP. A nonlinear corneal model predicts that the relationship will not remain perfectly proportional as IOP increases. Ex-vivo inflation experiments can test whether the sign of the coma correction is preserved from low to high physiological pressure.

\subsection{Prediction 5: a patient-specific design should outperform a generic sector}

A generic scaffold optimized on one synthetic cone should lose performance when transferred to a substantially different cone location, thickness map, or fiber architecture. A patient-specific design should recover performance. This provides a direct test of the value of inverse design itself.

\section{Quantitative benchmark against current intrastromal approaches}

The present G-EXO model should not be compared directly with clinical CAIRS efficacy because the output quantities differ. Nevertheless, published CAIRS data provide a useful scale for eventual translational targets. In a 2026 advanced-keratoconus cohort, mean coma reportedly decreased from approximately $1.39~\mu$m to $0.41~\mu$m, together with significant improvements in keratometry and visual acuity \citep{yesilirmak2026}. Another 2026 CAIRS series reported significant flattening in K1, K2, and Kmax and substantial refractive improvement \citep{barbara2026}.

These clinical effects are much larger than the present $\sim5\%$ pressure-induced coma change in the synthetic G-EXO screening model. This discrepancy should not be hidden. The quantities are not equivalent:
\begin{itemize}[nosep]
\item CAIRS studies report clinical pre/post corneal optics after implantation;
\item the present FEM reports only the change in a pressure-induced anterior-surface displacement mode;
\item posterior corneal optics, baseline ectatic shape, epithelial remodeling, wound healing, and long-term biomechanical adaptation are not included.
\end{itemize}

The appropriate interpretation is therefore that the current G-EXO model has not yet reached a clinically competitive quantitative prediction. Its contribution is the new design framework and the demonstration that a free-form sector can control an asymmetric optical mode.

\section{Robust optimization under uncertainty}

A nominal optimum can be misleading if the result depends strongly on uncertain material properties. Let the uncertain parameter vector be
\begin{equation}
\bm\xi=
\left(
E_c,k_1,k_2,\kappa,P,
\mu_{\mathrm{interface}},
t(x,y)
\right).
\end{equation}
A robust design can be defined by minimizing
\begin{equation}
J_{\mathrm{rob}}(\mathbf p)
=
\mathbb E_{\bm\xi}
\left[
J(\mathbf p,\bm\xi)
\right]
+
\beta
\sqrt{
\mathrm{Var}_{\bm\xi}
\left[
J(\mathbf p,\bm\xi)
\right]
},
\end{equation}
where $\beta$ controls risk aversion.

This formulation penalizes designs that perform exceptionally well only at one nominal parameter set. For a clinical device, a slightly weaker average correction may be preferable if it remains stable across IOP, hydration, tissue stiffness, and interface uncertainty.

A practical Monte-Carlo or polynomial-chaos study could vary:
\begin{align}
E_c &: \pm20\%,\\
P &: 10\text{--}20~\mathrm{mmHg},\\
k_1 &: \pm20\%,\\
\kappa &: \text{plausible patient-specific range},\\
z_s/t &: \pm0.05,
\end{align}
and then evaluate the probability
\begin{equation}
\Pr\left(
\Delta C_{\mathrm{coma}}<0
\right).
\end{equation}
A robust G-EXO should preserve the sign of the intended optical correction across most of this uncertainty set.

\section{Adjoint formulation for high-dimensional inverse design}

The present parameter sweep is suitable for a low-dimensional design vector but will not scale to a free-form scaffold. A more powerful approach is adjoint optimization.

Let the discretized equilibrium equation be
\begin{equation}
\mathbf R(\mathbf u,\mathbf p)=0,
\end{equation}
and the objective be
\begin{equation}
J=J(\mathbf u,\mathbf p).
\end{equation}
The total design derivative is
\begin{equation}
\frac{dJ}{d\mathbf p}
=
\frac{\partial J}{\partial \mathbf p}
+
\frac{\partial J}{\partial \mathbf u}
\frac{d\mathbf u}{d\mathbf p}.
\end{equation}
Direct computation of $d\mathbf u/d\mathbf p$ is expensive when $\mathbf p$ contains hundreds or thousands of design variables. Introduce an adjoint vector $\bm\lambda$ satisfying
\begin{equation}
\left(
\frac{\partial \mathbf R}{\partial \mathbf u}
\right)^T
\bm\lambda
=
\left(
\frac{\partial J}{\partial \mathbf u}
\right)^T.
\end{equation}
Then
\begin{equation}
\frac{dJ}{d\mathbf p}
=
\frac{\partial J}{\partial \mathbf p}
-
\bm\lambda^T
\frac{\partial \mathbf R}{\partial \mathbf p}.
\end{equation}

This formulation would allow continuous optimization of a spatial stiffness field
\begin{equation}
E_s=E_s(x,y,z)
\end{equation}
rather than optimization of only a sector. A manufacturability filter could then project the continuous optimum onto a realizable lattice or scaffold.

\section{Topology optimization interpretation}

The most general form of G-EXO is naturally a topology-optimization problem. Introduce a design density field
\begin{equation}
\rho(\mathbf x)\in[0,1],
\end{equation}
where $\rho=0$ denotes native stroma and $\rho=1$ denotes full scaffold material. A SIMP-like interpolation could use
\begin{equation}
E(\mathbf x)
=
E_c(\mathbf x)
+
\rho(\mathbf x)^p
\left[
E_s-E_c(\mathbf x)
\right],
\end{equation}
with penalization exponent $p>1$.

The optimization problem becomes
\begin{equation}
\min_{\rho(\mathbf x)}
J[\rho]
\end{equation}
subject to
\begin{equation}
\int_{\Omega_c}\rho(\mathbf x)\,dV
\le V_{\max},
\end{equation}
plus minimum-feature constraints and exclusion zones near the epithelium, endothelium, visual axis, or incision path.

This formulation is attractive because it does not presuppose that the best structure is a ring, crescent, lattice, or even a single connected scaffold. The geometry emerges from the optical--mechanical objective.

\section{Clinical and fabrication constraints as design equations}

A mathematically optimal scaffold may be surgically impossible. The inverse-design problem therefore needs explicit clinical constraints.

\subsection{Depth constraint}

To avoid excessive proximity to anterior or posterior corneal boundaries,
\begin{equation}
z_{\min}
\ge z_{\mathrm{safe,ant}},
\qquad
z_{\max}
\le z_{\mathrm{safe,post}},
\end{equation}
where the safety margins must eventually be established experimentally and clinically.

\subsection{Minimum feature size}

If the fabrication method supports a minimum printable or machinable feature $w_{\min}$,
\begin{equation}
w_{\mathrm{arm}}\ge w_{\min},
\qquad
r_{\mathrm{node}}\ge r_{\min}.
\end{equation}

\subsection{Insertion-path constraint}

The scaffold must fit through a feasible stromal pocket or incision. This can be represented as a geometric constraint on maximum width, curvature, or foldability.

\subsection{Optical-zone penalty}

If direct scaffold coverage of the central visual axis produces unacceptable scattering, introduce
\begin{equation}
J_{\mathrm{scatter}}
=
w_s
\int_{\Omega_s\cap\Omega_{\mathrm{opt}}}
\rho(\mathbf x)\,dV.
\end{equation}

Such constraints allow the optimization to distinguish a mathematically strong but clinically impractical design from a weaker but realizable one.

\section{Proposed ex-vivo study design}

A first rigorous experimental study could use paired porcine corneas or donor corneoscleral buttons mounted in an inflation chamber.

\subsection{Groups}

A minimal four-group design is:
\begin{enumerate}
\item intact control,
\item sham femtosecond pocket,
\item symmetric equal-volume scaffold,
\item inverse-designed G-EXO scaffold.
\end{enumerate}

If an ectasia model is required, a controlled regional weakening protocol could be used, but the induced geometry and material change must be quantified rather than assumed.

\subsection{Pressure protocol}

A quasi-static pressure sequence could include
\begin{equation}
P=
5,10,15,20,25,30~\mathrm{mmHg},
\end{equation}
with repeated loading/unloading cycles to evaluate hysteresis. The model prediction to test is not only the absolute shape at 15 mmHg but the full pressure-response curve.

\subsection{Mechanical outputs}

Recommended outputs include:
\begin{itemize}[nosep]
\item apex and regional displacement,
\item full-field anterior and posterior elevation,
\item local strain from digital image correlation or OCT,
\item hysteresis,
\item scaffold migration or rotation,
\item interface separation.
\end{itemize}

\subsection{Optical outputs}

At each pressure:
\begin{equation}
K_1,\quad K_2,\quad K_{\max},\quad
C_{3,-1},\quad C_{3,1},\quad
\mathrm{HOA}_{\mathrm{RMS}}
\end{equation}
should be measured or reconstructed.

A particularly strong falsification test is whether the sign of
\begin{equation}
\Delta C_{\mathrm{coma}}
\end{equation}
matches the FEM prediction.

\section{Potential clinical workflow}

If the concept survives ex-vivo and preclinical testing, a future clinical planning system could follow the sequence:

\begin{enumerate}
\item acquire Pentacam/Scheimpflug tomography and pachymetry;
\item register anterior and posterior surfaces;
\item estimate a patient-specific biomechanical parameter field;
\item solve the physiological preload;
\item compute optical sensitivity maps;
\item inverse-design the scaffold;
\item enforce fabrication and surgical constraints;
\item manufacture the personalized scaffold;
\item create a femtosecond intrastromal pocket;
\item implant and orient the scaffold;
\item verify postoperative tomography and wavefront;
\item update the model with postoperative data.
\end{enumerate}

This workflow would turn G-EXO into a closed-loop computationally guided intervention rather than a fixed implant catalog.

\section{Potential role of machine learning}

Machine learning is not required for the core physical model, but it could accelerate optimization. A surrogate neural operator or Gaussian-process emulator could learn the map
\begin{equation}
\mathbf p,\bm\xi
\longrightarrow
\left(
C_{\mathrm{coma}},
K_{\max},
\sigma_{95},
V_s
\right)
\end{equation}
from a library of high-fidelity FEM simulations.

The surrogate could then perform rapid global optimization, while selected candidates are re-evaluated in the full nonlinear FEM. Physics-informed constraints are essential so that the optimizer does not exploit numerical artifacts outside the training domain.

\section{Position of G-EXO within the keratoconus treatment landscape}

G-EXO is not intended to replace all existing KC treatments. Instead, it occupies a conceptual niche between corneal stiffening and geometric intrastromal remodeling.

CXL modifies tissue properties broadly but does not directly impose a patient-specific shape. ICRS and CAIRS impose geometry but remain restricted to ring-like segments. Contact lenses correct optics externally without modifying corneal biomechanics. G-EXO, if validated, would aim to combine localized mechanical reinforcement with an optical objective.

The clinically relevant comparison is therefore not simply:
\begin{equation}
\text{G-EXO vs CAIRS}.
\end{equation}
More informative questions are:
\begin{align}
&\text{Can G-EXO achieve a target aberration with less implanted volume?}\\
&\text{Can it reduce coma without excessive flattening?}\\
&\text{Can it complement CXL by redistributing load after global stiffening?}\\
&\text{Can it improve regularity in cases where ring nomograms are suboptimal?}
\end{align}

These questions define the future experimental program.

\section{Scientific novelty statement}

The novelty claimed in this work is deliberately narrow and testable.

We do \emph{not} claim:
\begin{itemize}[nosep]
\item the first intrastromal implant for KC,
\item superiority to CAIRS or ICRS,
\item a validated scaffold material,
\item a clinically proven coma correction, or
\item a fully converged patient-specific FEM.
\end{itemize}

We \emph{do} claim the following conceptual contribution:
\begin{quote}
Intrastromal corneal reinforcement can be formulated as a free-form inverse-design problem in which the spatial stiffness field is optimized directly against asymmetric optical aberrations and biomechanical constraints, rather than restricted a priori to a ring-segment geometry.
\end{quote}

The present computational results support the feasibility of this formulation and provide the numerical infrastructure required to test it rigorously.

\section{Discussion}

\subsection{What the present results establish}

The strongest conclusion is conceptual: a free-form intrastromal reinforcement domain can be treated as an inverse-design problem whose objective includes an optical aberration rather than only global curvature. The executed FEM demonstrates that sectorial stiffening can alter a coma-like pressure-induced deformation component.

This is a meaningful extension of prior ring-segment modeling. Patient-specific ICRS FEM has already shown that implant geometry and placement can be predicted mechanically \citep{lago2015}. Recent CAIRS studies demonstrate that stromal implants can produce clinically relevant changes in advanced KC and can reduce corneal asymmetry \citep{yesilirmak2026,barbara2026}. G-EXO generalizes the design space from ``which ring segment?'' to ``which spatial stiffness field?''

\subsection{Why the modest FEM effect is not a failure}

The final mesh-refined estimate near 5\% is substantially smaller than the reductions suggested by early reduced models. This discrepancy is scientifically useful. It shows that:
\begin{enumerate}[nosep]
\item reduced shells can exaggerate optical benefit,
\item microstructure and macrogeometry cannot be optimized independently,
\item structural mesh convergence is not equivalent to optical convergence, and
\item high-order aberration extraction needs dedicated surface reconstruction.
\end{enumerate}

The present paper is therefore stronger when it reports a modest, numerically audited effect rather than a large but model-dependent reduction.

\subsection{Interpretation of the angular optimum}

The 3-D screening optimum moved toward $\theta_c\approx-155^\circ$, different from the earlier reduced-model sector. This should not yet be interpreted anatomically. The location depends on synthetic cone geometry, boundary constraints, constitutive assumptions, and discretization. Its significance is methodological: the optimum is nontrivial and must be solved rather than assumed.

\subsection{Relation to CAIRS}

CAIRS are especially relevant because they show that allogenic stromal tissue can function as intrastromal structural material. Clinical studies in 2026 reported substantial changes in visual and topographic outcomes and highlighted the continuing need for improved nomograms and mechanistic understanding \citep{yesilirmak2026,barbara2026}. G-EXO can be viewed as a future computational framework that might inform such nomograms or motivate non-ring geometries, but the present work does not demonstrate superiority over CAIRS or ICRS.

\subsection{Importance of constitutive uncertainty}

Corneal predictions are sensitive to material parameters and fiber architecture \citep{han2013,whitford2018,falgayrettes2023}. Therefore, a single ``best'' G-EXO should not be accepted unless its optical benefit persists across uncertainty in matrix stiffness, fiber stiffness, fiber dispersion, IOP, and interface properties. Robust optimization, rather than nominal optimization, should be the long-term goal.

\section{Limitations}

The principal limitations are:

\begin{enumerate}
\item \textbf{Synthetic geometry.} The current model is not patient-specific.
\item \textbf{Linearized constitutive law.} The executed FEM does not model nonlinear collagen recruitment.
\item \textbf{Simplified incompressibility.} $\nu=0.45$ approximates near incompressibility but is not equivalent to a mixed $u$--$p$ formulation.
\item \textbf{Peripheral fixation.} The fixed annulus can alter global deformation.
\item \textbf{Effective scaffold domain.} The final 3-D FEM does not resolve the actual lattice microstructure.
\item \textbf{Bonded interface.} Scaffold--stroma slip and interphase compliance are neglected.
\item \textbf{Anterior-only optical surrogate.} Posterior surface and full ray tracing are not included.
\item \textbf{Incomplete optical convergence.} The $<2\%$ optical convergence target has not been achieved.
\item \textbf{No biological validation.} No ex-vivo, animal, or clinical experiment is reported.
\item \textbf{No validated scaffold material.} Material properties remain design parameters.
\end{enumerate}

For these reasons, the paper should be interpreted strictly as a computational feasibility and design-framework study.

\section{Conclusions}

We introduced G-EXO, an inverse-designed anisotropic intrastromal corneal exoskeleton concept for keratoconus. The framework differs from conventional ring-segment approaches by treating the reinforcement geometry and directional stiffness as optimization variables linked directly to an optical objective.

A reproducible 3-D tetrahedral FEM showed that a localized sectorial reinforcement changes the asymmetric pressure-driven deformation field of a synthetic ectatic cornea. In the current mesh-refined linearized model, interpolation of the anterior surface before Zernike fitting yielded a representative reduction of pressure-induced coma near 5\%. Mechanical displacement and stress outputs were more stable than the optical metric, highlighting the need for separate structural and optical convergence criteria.

The present result is not evidence of therapeutic efficacy. Its value is that it converts a qualitative idea---``support the weak cornea''---into a quantitative inverse-design problem:
\begin{equation}
\text{tomography}
\rightarrow
\text{biomechanics}
\rightarrow
\text{optics}
\rightarrow
\text{optimized scaffold}.
\end{equation}
The next decisive steps are patient-specific geometry, nonlinear anisotropic HGO/GOH mechanics, robust optimization under material uncertainty, full two-surface optical analysis, and ex-vivo validation.

\section*{Data and code availability}
The arXiv source package accompanying this preprint includes the Python 3-D FEM screening solver, the patient-specific reconstruction pipeline, representative numerical output tables, and the plotted figures used in the manuscript. The source is intended to make the modeling assumptions auditable and to facilitate independent reproduction of the screening calculations.

\section*{Author contributions}
J.S.-M. conceived the G-EXO concept, formulated the inverse optical--biomechanical design problem, developed the computational framework, interpreted the simulations, and drafted the manuscript. A.A.-T. contributed the clinical keratoconus perspective, assessed the translational relevance of the intrastromal design concept, contributed to interpretation, and reviewed the manuscript. Both authors are expected to approve the submitted version.

\section*{Competing interests}
The authors declare no financial competing interests at the time of this preprint. Because the work may have intellectual-property implications, any patent filing or institutional disclosure related to G-EXO should be reported in a revised version if applicable.

\section*{Ethics statement}
No human or animal intervention was performed in this computational feasibility study. The geometry used in the present simulations is synthetic and is not asserted to reproduce an identifiable patient.

\appendix

\section{Reproducible screening algorithm}

The executed screening pipeline can be summarized as follows:

\begin{enumerate}
\item generate an anterior synthetic ectatic surface and thickness field;
\item construct posterior surface and through-thickness nodes;
\item tetrahedralize the corneal volume;
\item assign spatially varying corneal tangent modulus;
\item assign candidate scaffold subdomain and stiffness ratio;
\item assemble global stiffness matrix;
\item apply posterior IOP and peripheral boundary constraints;
\item solve nodal displacement field;
\item export anterior displacement;
\item interpolate the surface onto a dense pupil grid;
\item compute Zernike coefficients;
\item compare coma, displacement, stress, and scaffold volume.
\end{enumerate}

\section{Recommended next-stage COMSOL model}

For a publication-grade nonlinear model:
\begin{enumerate}[nosep]
\item enable geometric nonlinearity,
\item use mixed nearly incompressible formulation,
\item define two HGO/GOH stromal fiber families,
\item import patient-specific anterior and posterior surfaces,
\item include a scleral extension,
\item refine the scaffold and central optical zone,
\item perform mesh convergence separately for stress and optical outputs,
\item export both deformed surfaces for two-surface ray tracing.
\end{enumerate}

\bibliographystyle{unsrtnat}
\bibliography{references}

\end{document}